\newcommand{\beq}{\begin{equation}}
\newcommand{\eeq}{\end{equation}}
\newcommand{\beqns}{\begin{equation}}
\newcommand{\eeqns}{\end{equation}}
\newcommand{\beqar}{\begin{eqnarray}}
\newcommand{\bs}{\begin{eqnarray*}}
\newcommand{\eeqar}{\end{eqnarray}}
\newcommand{\es}{\end{eqnarray*}}
\newcommand{\beqml}{\begin{mathletters}}
\newcommand{\eeqml}{\end{mathletters}}
\newcommand{\x}{\mbox{\boldmath $x$}}
\newcommand{\0}{\mbox{\boldmath $x$}_0}

\newfont{\fancy}{msbm10 scaled\magstep1}

\documentstyle[aps,preprint]{revtex}
\begin{document}
\draft
\preprint{PURD-TH-93-02}

\title{Weak non Self-Averaging Behaviour\\
 for Diffusion in a Trapping Environment \\}
\author{Achille Giacometti }
\address{Department of Physics, Purdue University,
West Lafayette, IN,47907}
\author{Amos Maritan}
\address{Dipartimento di Fisica, Universita' di Padova \\
via Marzolo 8, 35131 Padova}
\date{\today}
\maketitle
\begin{abstract}
The statistics of equally weighted random paths (ideal polymer)
is studied in $2$ and $3$ dimensional percolating clusters.
This is equivalent to
diffusion in the presence of a trapping environment. The number of
$N$ step walks follows a log-normal distribution with a variance
growing asymptotically faster than the mean which leads to a weak
non self-averaging behaviour. Critical exponents associated
with the scaling of the two-points correlation function do not
obey standard scaling laws.
\end{abstract}
\vspace{5mm}
%
%%%%%%%%%%%%%%%%%%%%%%%%%%%%%%%%%%%%%%%%%%%%%%%%%%%%%%%%%%%%%%%%%%%%%%%%%%
%
%                                   TEXT
%
%%%%%%%%%%%%%%%%%%%%%%%%%%%%%%%%%%%%%%%%%%%%%%%%%%%%%%%%%%%%%%%%%%%%%%%%%%
%
\newpage
\narrowtext
Diffusion of independent particles in the presence of randomly
distributed traps has been studied for a variety of purposes.
Besides being a prototype of a disordered system which allows
for some analytical treatment \cite{HK,N}, it may describe
migration properties of excitons in mixed organic crystals \cite{W}
as well as diffusion-controlled reaction of diffusing particles
with immobile centers.

A much less trivial problem is the diffusion of independent particles
on a percolating cluster \cite{C} when the environment acts as a perfect trap.
Specifically let us consider a lattice where a given site acts as
a perfect trap for particle motion with probability $1-p$.
We will consider the diffusion problem on the incipient infinite cluster
at the percolation threshold, $p=p_c$, both in $2$ (square lattice)
and $3$ (cubic lattice) dimensions, where particles can jump only
between nearest neighbours sites and are absorbed upon hitting a trap.
Thus at variance with previous studies where diffusion occurs on
arbitrary (finite and infinite) clusters \cite{HK},
we will focus on the statistics of random paths having
common origin in the incipient infinite cluster.

This model differs from the usual kinetic random paths
(random walk) on the same structure in the fact that the weight
given to each path is (a constant) independent of the particular
path, and that the total probability is not conserved \cite{M,G}.
Thus this model is the exact equivalent of the well known self-avoiding
walk (SAW) problem on the same structure when self-avoidance
is negligible, and is also known as the ideal chain (IC) problem \cite{M}.

In view of the fact that the SAW problem on strongly correlated
disorder (such as the incipient infinite cluster)
appeared to be extremely hard to understand \cite{MN}, it is
worthwile to first understand the simpler problem of a single
polymer coupled with disorder but without excluded volume effects.

We will follow the method developed in a previous work \cite{G}.
The basic idea of this method is that the discretized version
of the Master equation for the probability
$P_{\0,\x}(N)$ of being at site $\x$ after $N$ steps, having started
from $\0$ on a lattice of coordination $z$,
can be expressed as a product of random matrices (transfer matrices)
containing the information about the percolating cluster ${\cal C}$.
The number of $N$ step walks in ${\cal C}$ with extrema $\0$,$\x$ is then
$C_{\0,\x}(N) = z^N P_{\0,\x}(N)$ and the total number
of walks with origin in $\0$ is $C_N=\sum_{\x}C_{\0,\x}(N)$, which
will play a role similar to the partition function.

Let ${\cal P}({\cal C})$ be the probability of ${\cal C}$.
The quenched average for the $q-$th moment of the end-to-end
distance of an $N$ step walk $w$ is defined as:

\beqar \label{r2q}
\overline{<R_N^q>} &=& \sum_{{\cal C}} {\cal P}({\cal C})
[\frac{1}{C_N} \sum_{w} R^q(w) \theta(w;{\cal C})]
\eeqar

where $\theta(w;{\cal C})=1$ if $w\subseteq {\cal C}$ and $0$ otherwise,
and $R(w)$ is the distance between the extrema of a walk $w$.
Similarly the probability that a surviving (not trapped)
particle is at site $\x$ after an $N$ step walk, can be defined by
\mbox{$P(\x,N) \equiv C_{\0,\x}(N) /C_N$}
(which is different from the quantity \mbox{$P_{\0,\x}(N)
\equiv C_{\0,\x}(N)/z^N$} appearing in the Master equation because
it is normalized to unity, while the latter is a non-conserved quantity).

It is important to notice that the disorder average was
performed only on configurations ${\cal C}$
which span the lattice at the percolation threshold
(that is the subset of the clusters which are infinite),
each configuration counted once in sampling.

Our best estimate for the exponent associated with the end-to-end distance
$\overline{<R_N^2>} \sim N^{2 \nu}$ is
\mbox{$\nu=0.58 \pm 0.03$} in \mbox{$d=2$} and
\mbox{$\nu=0.50 \pm 0.03$} in \mbox{$d=3$} (see Fig.\ \ref{fig1}).
These values were obtained as an average of
values calculated by  a regression procedure, Pade' techniques
and standard extrapolating methods. The errors are statistical.

The data employed were obtained by averaging $5$ different
sets consisting of $1000$ configurations each.
The number of sites in the configurations was typically
$15000$ for $d=2$ and $40000$ for $d=3$.  The maximum number of steps
was $N=1600$ and $N=600$ in $d=2,3$ respectively, well below the point
at which finite size effects become detectable.
In $d=3$ higher values of $N$ do not change the previous estimate
within our numerical errors.
It should be noted that there seems to be a crossover
between a subdiffusive regime ($\nu \leq 0.5$, for both $d=2$ and $d=3$),
to a regime where the system is superdiffusive ($\nu > 0.5$),
(see Fig.\ \ref{fig1}), starting around $N=100$.
In Fig. \ref{fig3} it is shown the local slope $\nu_N$ as a function of
$1/N$ in $d=2$, which gives support to the above value.

We also stress that the present problem is different from
another interesting model, the freely jointed
chain (with no excluded volume) in presence of random
obstacles \cite{WHC}, because our initial starting point
is anchored. In this latter case a stretched chain is to be
expected, as discussed in \cite{CB}, and Flory-Lifshits
arguments do not apply.

Scaling of the quenched average of the probability density
to be at site ${\bf R}$ at the discrete time $N$ has been numerically
investigated using the standard ansatz \cite{HB}:

\beqar \label{scaling r}
\overline{P({\bf R},N)}&=&\frac{1}{N^{\nu d_F}}
F( \frac{R}{N^{\nu}})
\eeqar

\noindent
where $F(x)$ is a universal function such that
\mbox{$F(x) \sim exp(- x^{\delta})$} for $x>>1$ and \mbox{$F(x)
\sim x^g$} for \mbox{$x \rightarrow 0$} and $d_F$ is the fractal
dimension of the infinite incipient cluster equal to $91/48 \sim 1.9 $
and $ \sim 2.5$ in $d=2,3$ respectively \cite{C}.
Note that we have normalized the probability so that
\mbox{$\int d^dR \;\; \rho(R) \overline{P({\bf R},N)}= 1$},
where the density $\rho(R)$ takes into account the fact that the
support of the measure is fractal.

The universal function $x^{d_F}F(x)$ can be obtained by looking
at the probability that a walk is a distance $R$ from the origin. Using
the above values of $\nu$, a rather good collapse of the data
is found both in $d=2$ and $d=3$.
In Fig.\ \ref{fig4} the collapse of the data is shown in $d=2$ for
$N=400,800,1200$ and $1600$.  The best fit for is obtained
for $\delta=1.60 \pm 0.03$, which is not consistent with
the scaling relation \cite{F} \mbox{$\delta=
(1- \nu)^{-1} = 2.38 \pm 0.17$} taking for $\nu$ the
value previously given.
For small values of the argument we could fit
\mbox{$x^{d_F}F(x) \sim x^{d_F+ g}$}, with
\mbox{$g=-0.29 \pm 0.03$}.

Consistent values of $g$ and $\delta$ have been obtained by calculating
dimensionless ratios of $q-$th moments of the end-to-end
distance (see eq.\ (\ref{r2q})) and comparing them with the
prediction coming from eq.\ (\ref{scaling r}) and the assumed form
of the universal function $F$.

Results of almost the same quality have been obtained in
$d=3$ where we found $\delta=1.61 \pm 0.05$
and $g=-0.33 \pm 0.07$. The value of $\delta$ obtained from
the relation $(1-\nu)^{-1}$, would be $\delta=2.00 \pm 0.12$
which again is not compatible with the above numerical values.

{}From eq.\ (\ref{scaling r}) the quenched average of the return
probability behaves asymptotically as \mbox{$
\overline{P(\0,N)} \stackrel{N>>1}{\sim} N^{ - \frac{\tilde{d}}{2}}$}.
Here $\tilde{d}$ is given by \mbox{$\tilde{d}/2=(d_F+g) \nu$}, which
generalizes the Alexander Orbach relation \cite{AO}, and reduces to it
when $g=0$. A linear fit to the data shown in Fig.\ \ref{fig2} gives
\mbox{$\tilde{d}/2=0.94 \pm 0.06$}. This is in perfect accord
with the value \mbox{$(d_F+g) \nu=0.93 \pm 0.02$} obtained by using
the values of $\nu$ and $g$ previously calculated.

The values shown are derived from an average over 8
different sets with $1000$ of configurations each.
In $d=3$ we also find a value \mbox{$\tilde{d}/2=1.02 \pm 0.07$}
from the best fit, while \mbox{$(d_F+g) \nu=1.08 \pm 0.04$},
which is again compatible.

The exponent $g$ is a measure of how favorable the conditions are
for the walk to return to the origin. Unlike the SAW on a Euclidean lattice
where (\mbox{$g=(\gamma-1)/\nu>0$} \cite{des C})
and unlike the case of kinetically weighted paths on any
(disordered or not) structure where $g=0$ \cite{AO},
the IC has $g<0$ i.e. it is quite probable the return to the
starting point \cite{note3}.

In order better understand the role of large statistical fluctuations
\cite{D}, we evaluated numerically the probability density $P(C,N)$ for
the distribution of $C_N$ over different realizations of
disorder (analogous results have been obtained for the number of $N$ step
returning walks \mbox{$C_{\0,\0}(N)$}).
Figure\ \ref{fig5} shows the function \mbox{$P(ln C)\equiv C P(C,N)$}  for
$N=400,800,1200,1600$ for the case $d=2$ where $C\equiv C_N$.
They are  very well fitted by a log-normal distribution
(solid line), of the form:

\beqar \label{log-normal}
P(C,N)&=&\frac{1}{C\sqrt{2 \pi \sigma_N^2}} exp[\frac{-(ln C -\lambda_N)^2}
{2 \sigma_N^2}]
\eeqar

\noindent
where the mean $\lambda_N$, as well as the variance $\sigma_N^2$, depend
on $N$. In terms of the scaled variables \mbox{$(ln C-\lambda_N)/
\sqrt{2 \sigma_N^2}$} the values of \mbox{$\sqrt{2 \pi \sigma_N^2}
P(C,N)$}, nicely collapse onto a
single universal curve  (not shown)\cite{note6}.

It is easy to relate this result to the moments of the distribution.
Indeed from eq.\ (\ref{log-normal}), the logarithmic moment is
$Z_0(N)\equiv \overline{ln C_N} = \lambda_N$.

The direct numerical computation of the logarithmic moment
gives, for $N>>1$:

\beqar \label{log moment}
Z_0(N)&=& N ln \mu - \alpha N^{\psi}
\eeqar

with $\mu=3.76 \pm 0.02, \psi=0.80 \pm 0.01.$
and $\alpha=0.52 \pm 0.01$, in $d=2$ \cite{Ma}.
These values were determined by fitting the tail of the quantity
\mbox{$ln[Z_0(N+\Delta N)/Z_0(N)]$} on data
obtained by averaging over 8 different sets
of $1000$ different configurations each.
Almost identical values are found by making the linearization
transformation $\Delta N/N \rightarrow (\Delta N/N)^{1-\psi}$
which gives a straight line when $\psi=0.8$.
These values can be checked against the $N$ dependence obtained
for $\lambda_N$ assuming the log-normal distribution
eq.\ (\ref{log-normal}).
The values obtained for the above parameters are:
$\mu=3.76 \pm 0.01, \alpha=0.52 \pm 0.01, \psi=0.79 \pm 0.01$,
which is in very good agreement with the previous estimates.

In the three dimensions we found $\psi=0.85 \pm 0.01$ directly
and $\psi=0.85 \pm 0.03$, from the fit of $\lambda_N$ \cite{Ma}.
The other parameters were also compatible.

The variance $\sigma_N^2$ is directly related to the free energy fluctuations
since $\sigma_N^2= \overline{(ln C_N)^2}-(\overline{ln C_N})^2
\sim N^{2 \chi}$ where $\chi$ is a new exponent.
{}From the fit of the log-normal distribution in $d=2$ we obtain
$\chi =0.66 \pm 0.01$, which we checked against a direct computation
of the logarithmic moments, giving $\chi=0.68 \pm 0.01$.
In $d=3$ we found $\chi =0.72 \pm 0.01$ directly
and $\chi=0.70 \pm 0.01$ from the log-normal distribution.
Consistent values are obtained from extrapolation methods
and from the Pade' analysis.
Note that the exact inequality $(1-\chi)/\nu \leq d/2$ \cite{CCFS}
is always satisfied.

The fact that the values for the variance and the mean are
such that \mbox{$\sigma_N^2>>\lambda_N$}, for \mbox{$N>>1$}, is not surprising,
since there is no equivalent of the central limit theorem
for random multiplicative processes \cite{R}.
We shall refer to the condition $\sigma_N^q>>\lambda_N \; (q >1)$
but $\sigma_N<<\lambda_N$ as
{\it weak non-self-averaging behaviour}.
The averages are dominated by rare events
with large values and therefore do not represent
well the asymptotic behaviour of the system \cite{note2}.
In this respect, a direct evaluation of the probability distribution
is essential in order to understand the overall asymptotic behaviour
\cite{D}.

In the evaluation of the non-logarithmic moments like $\overline{C_N^q}$
using eq.\ (\ref{log-normal}), it is necessary to specify the upper
limit of integration i.e. $C_N \leq z^N$ (the lower limit is of course
$C_N \geq 1$). This stems from the fact that
the tail of the distribution does not decay
fast enough and thus violates Carleman's criterion \cite{O}.

It is easily shown that the upper and lower
cut-offs have no effect on the results for the moments
of $ ln C_N$ whose distribution is a sharply peaked
Gaussian for \mbox{$N>>1$}.

A straightforward calculation in the large $N$ limit gives

\beqar \label{final}
Z_q(N)&\equiv& \overline{C_N^q}^{1/q}=z^N exp[ - A N^{\phi}]
\eeqar
where $A=\frac{1}{2 q} [ ln(z/ \mu)]^2$ and $\phi=2(1-\chi)$,
leading to a survival probability $P_S(N)=\overline{C_N/z^N}
\sim exp[-A N^{\phi}] \;\;\;(N>>1)$.

Were the average unrestricted, i.e.
taken over all clusters (finite and infinite), then a rigorous
result for the asymptotic behaviour of the survival probability
in a disordered d-dimensional lattice in which perfect absorbing
traps are present (with probability $1-p$) \cite{DV,GP} would give
an exponent $\phi=d/(d+2)$.

Some attempts to identify when this asymptotic limit sets in, have
been unsuccessful \cite{HDKW} due to the large fluctuations
present, presumably of the same kind as those reported here.

No rigorous results are known, for the present case.
However a straightforward extension of the heuristic
argument previously applied to the case of unrestricted averages
\cite{GP} would lead to $\phi=d_F \nu/(d_F \nu+1)$ and
\mbox{$\chi= \frac{d_F \nu+2}{2 d_F \nu +2}$}.
With our values for $\nu$ this would yield $\chi=0.74 \pm 0.02$
in $d=2$ and $\chi=0.72 \pm 0.01$ in $d=3$, which is
satisfactory only in $d=3$.

In conclusion, we presented a detailed investigation
of the properties for the statistic of equally weighted random
paths on a percolating cluster both in 2 (square lattice) and 3
(simple cubic) dimensions.
Besides describing diffusion occurring only on the
incipient infinite cluster when the rest of the environment acts
as a perfect trap, this model corresponds to the usual self-avoiding
walk on the same structure, with the self-avoidance turned off.
The coupling between disorder and self-avoidance seems
indeed responsible for the lack of a definite plausible scenario
for the intriguing problem of the SAW on percolating clusters
\cite{MN}.
The main aim of the present work was to clarify the effect of
the fluctuations and is expressed in eqns.\ (\ref{log-normal}),
(\ref{log moment}) and (\ref{final}).
We also computed the quenched averaged end-to-end distances and
return probability in two and three dimensions.
A generalization of the relation between spectral
dimension and the fractal dimension of the lattice was proposed.
A proper methodology to deal with anomalies in the log-normal
distributions, which might be useful in other fields, was
also described.

We are grateful to Hisao Nakanishi for his fruitful comments
and suggestions.
We also thank Stuart Burnett and Greg Follis for useful discussions
related to the analysis of the data and to the Pade' technique.
%%%%%%%%%%%%%%%%%%%%%%%%%%%%%%%%%%%%%%%%%%%%%%%%%%%%%%%%%%%%%%%%%%%%%%%%%
%
%                               BIBLIOGRAPHY
%
%%%%%%%%%%%%%%%%%%%%%%%%%%%%%%%%%%%%%%%%%%%%%%%%%%%%%%%%%%%%%%%%%%%%%%%%%

%\newpage
%\begin{thebibliography}{99}

%%%%%%%%%%%%%%%%%%%%%%%%%%%%%%%%%%%%%%%%%%%%%%%%%%%%%%%%%%%%%%%%%%%%%%%%%
%
%                              FIGURES
%
%%%%%%%%%%%%%%%%%%%%%%%%%%%%%%%%%%%%%%%%%%%%%%%%%%%%%%%%%%%%%%%%%%%%%%%%%
%
%\end{description}
%
%\newpage
\begin{figure}
\caption{log-log plot for the quenched averaged end-to-end
distance $R_2(N) \equiv \overline{<R_N^2>}$ for $d=2 \;\; (\bigcirc)$
and $d=3 \;\; (\bigtriangleup)$. The solid line has a slope
of $1.16$ while the dotted line has $1.00$ for $d=2,3$ respectively.}
\label{fig1}
\end{figure}

\begin{figure}
\caption{log-log plot for the quantity $S(N)\equiv
1/\overline{P(\0,N)} \sim N^{\tilde{d}/2}$ for $d=2 \;\; (\bigcirc)$
and $d=3 \;\; (\bigtriangleup)$. The solid line has a slope
of $0.94$ while the dotted line has $1.02$
for $d=2,3$ respectively.}
\label{fig2}
\end{figure}

\begin{figure}
\caption{Local slope for the two exponents $\nu$ and $\tilde{d}/4$
as function of $1/N$ in the case $d=2$. Also shown by an arrow are the values
obtained from the Pad\'{e} analysis.}
\label{fig3}
\end{figure}

\begin{figure}
\caption{Plot of the universal function  $x^{d_F}F(x)$
versus the dimensionless quantity $x=R/N^{\nu}$ , in
$d=2$. The data shown are an average over five points.
The solid line is the best fit result which gives $\delta=1.60 \pm 0.03$
and $g=-0.29 \pm 0.03$.}
\label{fig4}
\end{figure}

%\figure{log-log plot of the return probability $P_0(N)\equiv \overline
%{P(\0,N)} \sim N^{-\tilde{d}/2}$ for $d=2 \;\;(\bigcirc)
%$ and $d=3 \;\;(\triangle)$. The best fit is obtained for
%$\tilde{d}/2=0.94 \pm 0.01 $ and $ 1.02 \pm 0.07$ for
%$d=2$ and $d=3$ respectively. The dotted line has a slope 1.00.
%\label{fig3}}

\begin{figure}
\caption{Calculated distribution $P(lnC)$.  The solid lines are the best
fit results derived from equation\ (\protect\ref{log-normal}).
The points shown are raw data without any smoothing procedure,
and with a normalization factor of $2.8^N$ for $C$.}
\label{fig5}
\end{figure}


\begin{references}
\bibitem{note6} We numerically checked the gaussian character of
$P(ln C)$ up to the $6-$cumulant.
\bibitem{note2} In ref. \cite{G} it was argued, on a basis
of a smaller statistic, that there was no correction
to the leading exponential term, and that all the moments
had a simple exponential growth $\mu^N$ ($\mu <z$).
In view of the results presented here, this conclusion
should be considered erroneous.
\end{references}
\end{document}